# Biophysical models of intrinsic homeostasis: Firing rates and beyond

Authors:

Nelson Niemeyer[1,2,3], Jan-Hendrik Schleimer[1,3], Susanne Schreiber[1,2,3]*

Affiliations:

[1]*Institute for Theoretical Biology, Humboldt-Universität zu Berlin, 10115 Berlin*

[2]*Einstein Center for Neurosciences Berlin, Charitéplatz 1, 10117, Berlin, Germany*

[3]*Bernstein Center for Computational Neuroscience, 10115 Berlin, Germany*

*Corresponding author: s.schreiber@hu-berlin.de

## Abstract

In view of ever-changing conditions both in the external world and in intrinsic brain states, maintaining the robustness of computations poses a challenge, adequate solutions to which we are only beginning to understand. At the level of cell-intrinsic properties, biophysical models of neurons permit one to identify relevant physiological substrates that can serve as regulators of neuronal excitability and to test how feedback loops can stabilize crucial variables such as long-term calcium levels and firing rates. Mathematical theory has also revealed a rich set of complementary computational properties arising from distinct cellular dynamics and even shaping processing at the network level. Here, we provide an overview over recently explored homeostatic mechanisms derived from biophysical models and hypothesize how multiple dynamical characteristics of cells, including their intrinsic neuronal excitability classes, can be stably controlled.

# Highlights

- Critical transitions in membrane voltage dynamics pose a particular challenge for homeostatic control of neural activity.

- Next to firing rates, homeostasis can target other computational properties such as the excitability class or burstiness.

- Beyond ion-channel expression, ionic concentration dynamics and energy consumption are likely to contribute to homeostasis.

# Introduction

As engineers know well, it does not suffice if a device operates in ideal conditions. Its functionality must also be preserved in a world full of expectable perturbations that threaten to compromise performance. The same aspiration holds for the design of nervous systems: robust processing needs to be guaranteed despite highly dynamic environments and inputs. Accordingly, nervous systems are thought to homeostatically regulate their activity, from the intrinsic dynamics of neurons to the level of synapses and networks, all aiming to preserve crucial characteristics of information processing while operating within a limited energy budget. Neurons have indeed been experimentally shown to stabilize their firing rate [1], for example by decreasing their excitability at the axon initial segment when faced with intense stimulation [2–4]. Complex mechanisms of synaptic scaling, modification of the inhibition-excitation balance, and neurogenesis ensure robustness at the network level [5] as reviewed by Zenke et al. [6] and Tien and Kerschensteiner [7].

A less intensively studied aspect is the homeostasis of computational characteristics arising from cell-intrinsic properties [8]. From a computational perspective, many such characteristics impact neuronal function and may hence form suitable targets for homeostatic regulation. In particular, these include not only firing rates, but also encoding properties such as signal filtering or the generation of specific spike patterns. Although experimental evidence in support of an intrinsic homeostasis of bursting [9,10]* and of synchronization properties [11] has been described, such 'computational homeostasis' beyond firing rates has not been extensively explored. The endeavor is complicated by the complexity of the potential target space. For example, neural tissue is energetically costly [12,13], and the homeostasis of energy metabolism may at times interfere with computational needs, despite the obvious link to strategies of efficient coding [14,15]. The relative relevance of potentially conflicting targets is likely to be determined by neuronal function. Thus computational variables need to be prioritized when essential information is at stake [16]*; energetic variables may rise in regulatory importance when pathophysiological conditions, such as ischemia, are likely to be encountered [17,18].

Even if the homeostatic targets are well-known, the question of a physiologically realistic control mechanism arises. In general, homeostatic targets can be stabilized by a variety of feedback loops, the most prominent mechanism relying on calcium-dependent changes in transcription rates of ion channel genes [19–21]*. Yet, also mechanisms beyond an activity-regulated transcription [22]**, including post-translational modifications of ion channels or (re)location of the axon initial segment [4,23], come into play. Less analyzed - though highly relevant for computational homeostasis - is the role of 'passive' biophysical mechanisms, such as those based on ionic concentrations or changes in morphology [24].

In the following, we discuss the homeostasis of neuronal dynamics and, in some cases, their relation to energy homeostasis. We zoom in on the opportunities and challenges posed by a regulation of action potential dynamics and neural excitability beyond pure firing rates. On the way, we encounter mechanisms that extend beyond adjusting ion channel densities and highlight how multiple mechanisms could interact to ensure robust homeostatic regulation.

## Beyond firing rates: qualitative switches in neuronal excitability

Despite their seemingly stereotyped, pulsatile nature, the basic units of neuronal signaling – the action potentials – are by no means all the same. Indeed, their computational properties strongly depend on the nonlinear dynamics of the spike-generating mechanism. These dynamics shape how information is processed [25,26] and influence how neurons behave in a network [27–30]. For example, they determine the average stimulus inducing a spike (spike-triggered average,) [31] and the shape of the f-I curve (firing rate to input relation). Therefore, for neurons, a homeostatic regulation of their intrinsic dynamics beyond the firing rate seems to be a useful endeavor.

Based on response characteristics, neurons have been grouped into different excitability classes [32,33]. These can also be identified in mathematical neuron models, where they are associated with different spike onset bifurcations (Box 1), that is, dynamically distinct transitions from rest to spiking [34‑36] with bifurcation-specific characteristics. Close to excitability switches (Box 1), small parameter changes can qualitatively impact a neuron's behavior. In principle, almost any physiological parameter can induce such a transition. In theoretical models of cortical neurons [37] and in experiments [38], cholinergic neuromodulation was shown to switch the spike-generating mechanism. So can the presence of autapses [39]* or of ethanol [40]. Mechanistically, those alterations that affect the timescales of gating variables differently in comparison to the timescale of the voltage variable are particularly prone to inducing switches [41]. The latter includes changes in temperature [28], capacitance [42]*, or external potassium concentration [43].

**Box 1:** Excitability switches and spike onset bifurcations.

The excitability class of a neuron summarizes its functional properties. This includes whether the neuron can fire at arbitrarily low frequencies [32], higher-order spike statistics [31,44], whether specific signaling frequencies are preferred (subthreshold resonance [45,46] and spike-time coding [31,47]), and the tendency of a neuron to synchronize in networks [27,48]. The excitability class is fundamentally related to the bifurcation(s) [35] that the voltage dynamics undergoes when switching from rest to spiking at the rheobase.

More precisely, spike onset constitutes a codimension-1 bifurcation (i.e. triggered by one parameter, such as an increase in input current), of which there are qualitatively different varieties, depicted as areas in the diagram of Figure 1. Here, spike onset is defined as the bifurcation that creates a stable limit cycle (tonic spiking) at the lowest possible input current. The accompanying functional properties are summarized in the table at the bottom of Figure 1. An excitability switch is, then, a qualitative change in the neuron's transition from rest to spiking. Mathematically, these excitability switches are referred to as codimension-2 bifurcations (corresponding to lines in Figure 1 and summarized in the upper table in Figure 1), because they require an additional parameter (beyond the input current) to be tuned.

The diagram in Fig. 1 is spanned by two fundamental parameters, membrane leak and capacitance, and sketches the overall topology of neighboring excitability classes and the switches between them (it is a new visualization taking [49,50] into account). A third parameter, the constant input current, is assumed to be set to the rheobase. The topology of excitability classes is shared by all conductance-based neuron models. Furthermore, up to distortions topologically similar diagrams are spanned by other parameters, such as temperature [28] or potassium reversal potential [43], provided these have a differential effect on the timescale of the current balance equation compared to that of the gating kinetics.

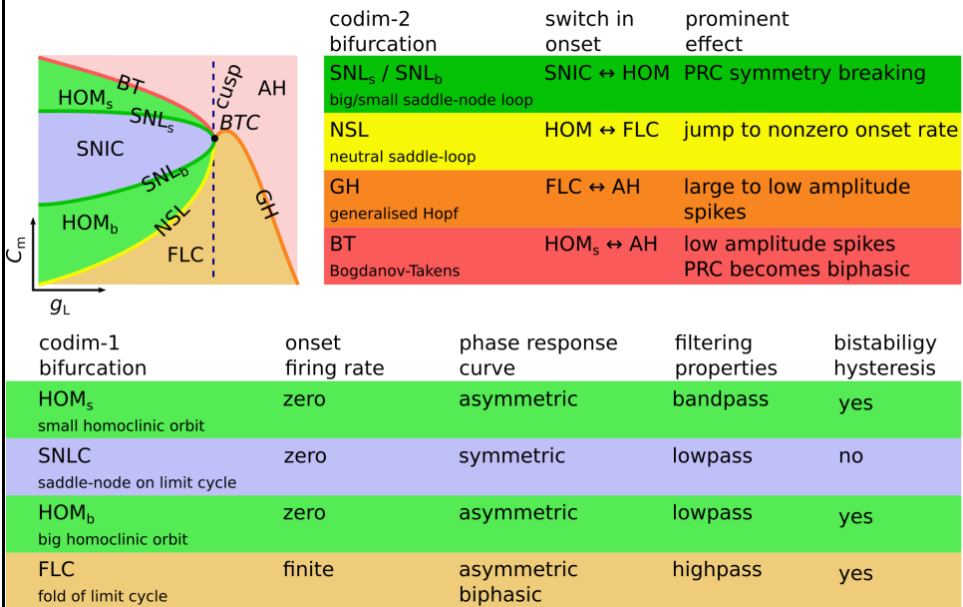

*Figure 1: Scheme based on [49,50] summarizing possible spike-onset bifurcations (areas, properties summarized in the bottom table) at rheobase and transitions between them (lines, properties summarized in the left table). The diagram is spanned by two-parameter axes, whereas the third parameter, the input current, is implicitly set to the rheobase, that is, the smallest constant input current that creates a large-amplitude, stable limit cycle (spike). AH denotes the Andronov-Hopf bifurcation, which is not included in the table of spike onset bifurcations as it creates either low amplitude (supercritical) or unstable (subcritical) limit cycles.*

## Homeostasis of excitability: a brief excursion into control theory

Already, a regulation of firing rates (and not the excitability class) is more challenging in the vicinity of critical switches. Established models for rate control rely on a smooth dependence between the regulating parameters (e.g. ionic conductances) and firing rates [19,51]**, which allows to linearize around set points. However, close to an excitability switch, the mapping between control parameters and homeostatic targets can jump discontinuously even for small parameter changes, which aggravates regulation. So far, it is hard to tell how much of a challenge this poses for established models of rate homeostasis because excitability switches encountered during regulation are typically not reported. Accordingly, future work should systematically compare regulation across switches and within an excitability class to clarify this issue.

Because of the tremendous importance of the excitability class on information processing and network behavior, we here propose that also the dynamical type, beyond the mere firing rate, should be under homeostatic control. In many ways, however, this type of homeostatic regulation is a more complicated venture than homeostasis of rates or $[Ca^{2+}]_i$ levels. Rates and concentrations are real numbers that can be ordered around set points, rendering up- and down-regulation meaningful distinctions [52] and allowing to clearly define feedback rules [53]. In contrast, the higher dimensional topology of bordering excitability classes is ruled by the fundamental unfolding of bifurcations in conductance-based neurons [34,36]. This poses novel questions and challenges:

1) Does a model augmented with a feedback regulatory mechanism preserve the fundamental bifurcation structure and hence exhibit excitability switches that are comparable to the original model? The answer depends on the control type. For example, integral controllers (relying on recent history) can be expected to affect the fundamental bifurcation structure, whereas derivative controllers (using local state predictions) cannot [54].

2) Could neurons keep a distance from excitability switches to stabilize the excitability class? To answer this question, the concept of "inverse bifurcation analysis" in mathematical models could be helpful, which has been developed to assess the closeness to switch-inducing bifurcations [55] and applied to systems biology [56].

3) Can control over the excitability class be achieved with feedback mechanisms? Although such control can be safely assumed for firing rates, this is not obvious for more complex excitability characteristics. Indeed, control of several switches between excitability classes found in neurons (Box 1) has been treated mathematically [57–59], including a control strategy to decide *a priori* which bifurcations are traversed at locations in parameter space where bifurcations branch and several options are available (such as at the so-called Bogdanov-Takens (BT) point in Box 1, where Hopf, saddle-node, and saddle-homoclinic orbit (HOM) bifurcations coincide) [57]. These ideas have been applied to the control of bifurcations in conductance-based neuron models [54,60]*, offering a mathematical control perspective yet currently leaving potential neural implementations unaddressed.

## Conductance-based mechanisms of excitability homeostasis

For firing rate homeostasis, physiological mechanisms of control have been explored and typically rely on negative feedback loops. Deviations from a target setpoint are reflected in a proxy signal, which in turn is translated into a biophysical parameter that drives the system back toward the target setpoint (Fig. 2a). A well-known feedback loop links intracellular calcium, $[Ca^{2+}]_i$, which correlates with recent neural activity, to changes in ion channel gene-transcription rates [19,53] (Fig. 2b). A timescale separation (between the fast spike-generating dynamics and the slower kinetics that govern the transcription rates) results in an integration of past spiking activity, enabling a regulation similar to an integral controller. The slower variable moves ion channel gene expression in a direction that allows the system to return to a target calcium level and firing rate [53].

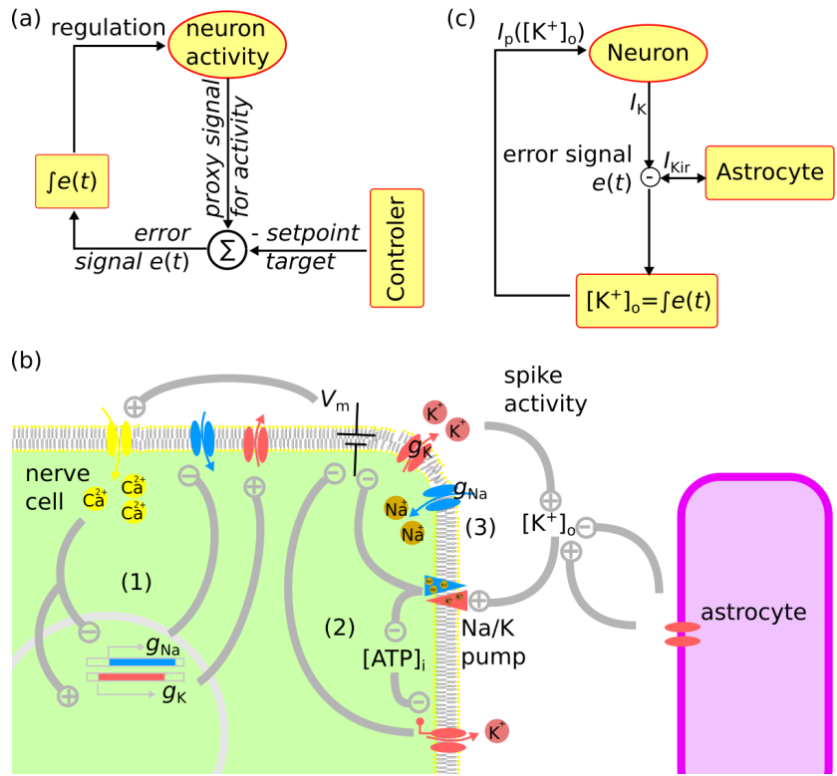

*Figure 2: Multiple feedback loops in a neuron. (a) Generic feedback controller. (b) Neuron in an extracellular bath indicating several (though not all) feedback mechanisms acting on spike activity. Among them are activity-dependent ion channel gene regulation (1), protective feedback loops involving hyperpolarizing potassium channels that are disinhibited by a drop in intracellular ATP (2) and hyperpolarizing pump currents owing to activity-induced pump load (3). (c) Formalized example of a feedback loop based on extracellular $K^+$ as a proxy for neuronal activity and astrocytic $K^+$ uptake as a controller: neuronal spiking activity leads to an increase in $[K^+]_o$ which drives electrogenic pump activity (this depends on the isoform [61]) and throttles neuronal activity. Astrocytes can also take up $[K^+]_o$ via $K_{ir}$ currents, thereby reducing the electrogenic pump activity and exercising control over the feedback.*

Yet the results of such a control are not unique; although fixed calcium targets are set, the resulting conductance parameters differ and, thus, identical calcium target levels (reached from different initial conditions) can be accompanied by largely different neuronal dynamics [19,51,62]**. This degeneracy has recently been addressed by adding post-transcriptional regulation [51]** to the modes of control, relying on additional interactions between channel precursors and removing the effects of initial conditions, therefore increasing robustness to perturbations. Moreover, recent experiments [63]* suggest that proteostatic mechanisms that respond to changes in channel expression independently of activity signals contribute to intrinsic homeostasis, potentially helping to specify a unique dynamical state.

However, the homeostasis question is still more complicated. Even if a unique combination of conductances could be specified from a mean $[Ca^{2+}]_i$ target, external influences (such as extracellular ionic concentrations or temperature [64,65]) may still distort the picture. For a homeostasis of the excitability class, this means that neurons cannot be certain of their dynamical type without means of measuring it. Such means could be envisaged if specific 'symptoms' of excitability [44], such as the emergence of interrupted firing owing to bistable states [66], were exploited. These characteristics based on spiking statistics would not be reflected in the averaged $[Ca^{2+}]_i$ concentration but in its higher-order statistics. Indeed, molecular $Ca^{2+}$ sensors with varying kinetics and optimal activation by distinct frequencies [67] are well-characterized and form suitable candidates for the regulation of $[Ca^{2+}]_i$ variance [68–71]*. Therefore, it is an attractive idea that neurons can use proxy signals over several timescales for the regulation of dynamical properties, choosing from several feedback loops available to neurons (Fig. 2b).

## The high degrees of freedom in biophysical neurons

Degeneracy, however, need not necessarily hinder intrinsic or computational homeostasis. Models adjusting channel densities revealed that degeneracy contributes to robustness of firing rate regulation [19] because perturbations in one channel density can be compensated through the other densities [53]. These models rely on a calcium-mediated coregulation, meaning that several densities are up- and down-regulated by a common master regulator. Indeed, sensitivity analysis can reveal effective regulatory directions in parameter space [52,72] when appropriate metrics (in this case, the difference to a target calcium level) can be defined [73,74]. Moreover, preservation of linear relationships between conductances can be robust as long as nonlinear effects on the target variable are moderate. Such models with coregulation successfully demonstrated robustness to many types of perturbations [53,62]. They are biologically plausible because all components are linked to the processes of protein expression.

More recently, it has even been proposed that the large number of adjustable parameters can benefit the regulation of multiple targets [21]*: even if for one target many parameter combinations form a solution, the number of possible solutions swiftly decreases if additional dynamical targets and constraints are considered. However, for a regulation of multiple targets, the integral control scheme of earlier models could not simply be extended because wind-ups, that is, unbound increases in maximal conductances, were easily produced. Instead, a modified rule based on proportional control (i.e. assuming a linear relationship between feedback strength and deviation to the target value) was used [21]* to meet a target firing rate and a target energy efficiency at the same time. The demonstration that simple homeostatic rules can balance two dynamical targets is a crucial step, yet the plausibility of biological implementations of feedback signals and proportional control seems less clear in this case.

## Energy homeostasis versus homeostasis by energy feedback

Irrespective of the excitability class, homeostasis of low firing rates (sparse spiking) has been motivated as an energy-saving strategy [75,76] because the individual spike is metabolically costly [13]. One may, therefore, wonder whether targeting firing rates and targeting energy consumption with homeostatic control are not the same. The following analogy with cruise control illustrates that the resulting behaviors are rather distinct: classic cruise control uses more fuel when driving uphill (keeping the velocity constant); an eco-cruise controller, on the contrary, permits a drop in velocity (keeping the fuel consumption constant). In analogy, homeostasis can preserve mean firing rates by investing as much energy as is required (an implicit assumption to all conductance-based neuron models with fixed ionic reversal potentials). Or, on the other hand, energy usage may be bounded, requiring firing rates to drop and rise, or neurons to even fall silent for periods of time (Fig. 3a). Neuronal means to achieve firing rate reductions for the sake of energy conservation exist and need not involve the direct regulation of ion channels, for example, when relying on the electrogenic nature of the $Na^+$-$K^+$-pump as an inhibitory feedback. In this case, extracellular potassium, $[K^+]_o$, serves as an integrated control variable similar to mRNA or $[Ca^{2+}]_i$, and firing rates are reduced [77,78] in the manner of an adaptation current (Fig. 2b,c).

Beyond firing rates, however, the interplay between neuronal dynamics (bifurcations) and metabolic supplies is intricate. For example, elevated neural activity increases extracellular potassium levels, $[K^+]_o$, and the altered $K^+$ reversal potential can induce a switch in the excitability class [41,43], see Box 1. Depending on pump densities, energy availability, or the level of potassium clearing by glia cells [79], there is yet another security net in place to keep neuronal firing rates in check on a longer timescale. Elevated $[K^+]_o$ also shifts the depolarization block [80] to lower input levels, effectively forcing the neuron to stop firing such that pumps can restore the ionic gradients (Fig. 3b). This built-in control mechanism would allow neurons to temporarily achieve high firing rates, while enabling the pumps to restore ionic gradients and, therefore, dynamical homeostasis during periods of silence.

A similar, yet more graded feedback mechanism relies on ATP-gated potassium channels which sense the cell's metabolic state [81–83]. Indeed, redundant feedback loops are a common motif in biology [84]. Nonetheless, most studies on neural homeostasis focus on an adjustment of channel densities, and alternative feedback loops are currently underrepresented. Especially when considering more complex dynamical properties or their combination, it may, however, be advantageous to rely on information from different proxies and multiple feedback loops.

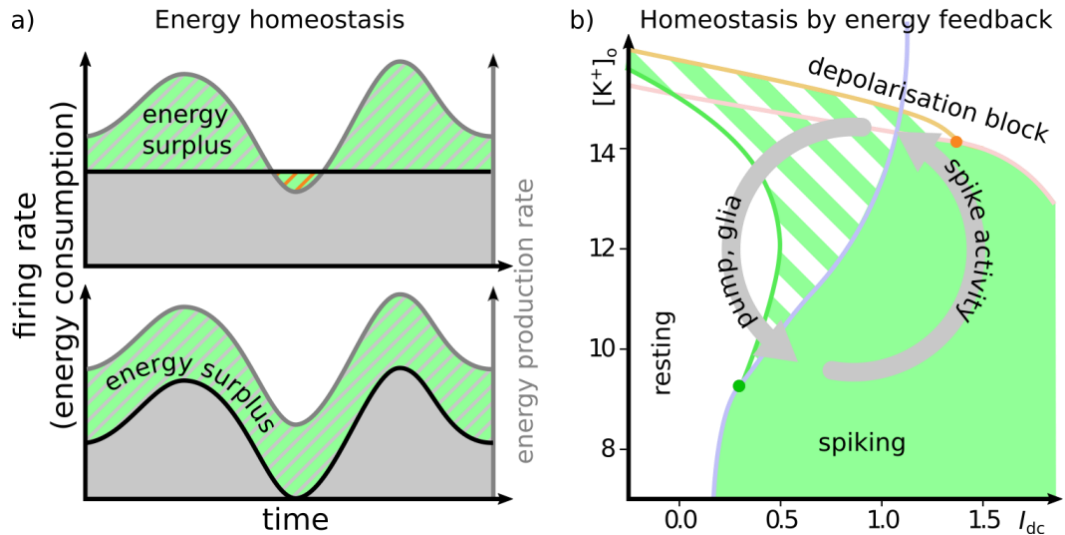

*Figure 3: Intrinsic homeostasis and energy (a) Owing to metabolite availability and other energy-consuming processes, cells have varying net energy production rates (gray curves). At the same time, the energy consumption of neural activity depends on the firing rate (black curves). The green area between these curves (energy surplus) represents the remaining energy available for other processes, such as protein synthesis. Top: homeostasis preserving firing rates can lead to a negative surplus and hence an energy crisis. Bottom: homeostasis preserving the energy surplus leads to firing rates that covary with energy production. (b) Energy homeostasis via the depolarization block. The spiking region (green) of a neuron and its bordering resting and depolarization block regions. Increased spiking leads to higher [K+]o levels, shifting the depolarization block to lower inputs till spiking becomes bistable (dashed region) and eventually terminates. The more neurons in the network fall silent, the lower the mean input to each neuron. During the silence periods, [K+]o concentration is reduced by K+-dependent pumps, re-enabling spiking (color code of bifurcations as in Box 1). The bifurcation diagram is based on a Wang-Buzsáki model with potassium concentration dynamics and an Na+-K+-pump.*

## Conclusions

'Computational homeostasis', that is, ensuring the robustness of function, is not only relevant at the level of synapses and networks but also at the level of cell-intrinsic dynamics. First, we highlighted homeostatic mechanisms that allow for a cellular regulation of the firing rate, showcasing possible strategies and challenges. In particular, firing rates do not define a unique point in the vast space of cellular parameters (degeneracy). Homeostatic regulation of rates can, therefore, be accompanied by changes in other computational properties, including the excitability class. More experimental work is required to uncover the exact extent to which these additional properties are homeostatically preserved. Second, we discussed how encoding properties (beyond the firing rate) could be robustly maintained, with a focus on excitability class and the underlying bifurcation structure. We pointed out that many physiological substrates (ranging from internal calcium levels to extracellular potassium concentrations and metabolic variables) can serve as regulatory proxies of homeostatic targets - circumstances that could benefit a parallel regulation of different dynamical properties.

## Acknowledgements

This project has received funding from the European Research Council (ERC) under the European Union's Horizon 2020 research and innovation program (grant agreement No 864243), the German Ministry of Education and Research (grant no. 01GQ1403), and the Einstein Foundation Berlin (grant no. EZ-2014-224).

## Outstanding papers

- [10]*: This study in the crab shows that elevations in extracellular potassium can transiently abolish neuronal activity. While the loss is presumed to arise from synaptic processes, a restoration of bursting firing patters is mediated by changes in intrinsic excitability of the pacemaker neurons on the timescale of minutes.

- [16]*: MSO cells in the auditory brain stem are crucial for the computation of sound source direction. This modelling study demonstrates that electrophysiological and morphological properties of MSO cells are optimized for the computation of coincidence detection. Where possible without functional impairment, however, the metabolic cost of processing is minimized too.

- [21]*: This article addresses how multiple dynamical targets (firing rate and energy per spike) can be regulated in parallel. Exploring the biophysical conductance space, it is shown how homeostatic rules can find a solution for both targets. In this case, degeneracy in channel types helps to achieve robust homeostasis.

- [22]**: This experimental paper shows that mechanisms beyond an activity-regulated transcription of ion channel genes can contribute to firing rate homeostasis. Firing rate homeostasis is tested by prolonged pharmacological stimulation in cultured neurons of the mouse. Firing rate adaptation prevails even if transcription factors that are involved in activity-regulated transcription are knocked out or, alternatively, all transcription is acutely blocked pharmacologically. After excluding synaptic scaling effects, the authors suggest that the observed firing rate homeostasis relies on an adaptation of cell-intrinsic excitability.

- [39]*: This paper shows that autapses in biophysical neuron models can create transitions between excitability classes. This finding adds a new feedback mechanism to the set of potential regulators of the neuronal excitability class.

- [42]*: This mathematical paper explores a specific switch between excitability classes that ubiquitously occurs in conductance-based neuron models and can be induced by changes in several physiological parameters, including capacitance and membrane leak and results in changes of information transfer as well as network synchronization. Whereas previously more attention had been given to the Bogdanov-Takens (BT) point, this paper shows significant changes in the qualitative behaviour of neuron models that arise via the particular switch at the saddle-node loop.

- [51]**: This theoretical paper adds post-transcriptional interactions to previous models of transcription regulation of channel genes for intrinsic homeostasis. These interactions create a strong attractor in the conductance space and render homeostatic regulation more robust to initial conditions and to perturbations. Transcription rates are believed to be noisy and the model extension plausibly explains how robust regulation can still be achieved.
- [60]*: This study shows that a differential controller is efficient in regulating the excitability class in a biophysical model neuron, which is interesting because most models of homeostasis use integral controllers.
- [63]*: In this fascinating study the homeostatic responses to two dynamically equivalent perturbations (knocking down or blocking the same channel conductance) are shown to involve distinct mechanisms. This suggests that activity-independent, proteostatic mechanisms act alongside activity-dependent mechanisms to regulate firing rates. One reason why this is interesting is that in degenerate conductance spaces proteostatic feedback could help to specify a subset from the many solutions that yield equivalent calcium levels.
- [71]*: This paper provides both a review as well as original work. The authors make a compelling case that for efficient information transfer, not only the mean firing rate but also its variance needs to be matched to the input. Models with two controllers are designed to regulate both of these properties. Conditions are described when these controllers work in tandem and when they produce a wind-up (runaway unphysiological values). The paper also suggests interesting experiments to distinguish between scenarios of single versus dual controllers for the molecular pathways that regulate synaptic conductance and intrinsic excitability.

# References and recommended reading

1. Turrigiano G, Abbott LF, Marder E: **Activity-dependent changes in the intrinsic properties of cultured neurons**. *Science* 1994, **264**:974–977.

2. Grubb MS, Burrone J: **Activity-dependent relocation of the axon initial segment fine-tunes neuronal excitability**. *Nature* 2010, **465**:1070–1074.

3. Kuba H, Oichi Y, Ohmori H: **Presynaptic activity regulates Na+ channel distribution at the axon initial segment**. *Nature* 2010, **465**:1075–1078.

4. Goethals S, Brette R: **Theoretical relation between axon initial segment geometry and excitability**. *eLife* 2020, **9**.

5. Gjorgjieva J, Evers JF, Eglen SJ: **Homeostatic Activity-Dependent Tuning of Recurrent Networks for Robust Propagation of Activity**. *J Neurosci* 2016, **36**:3722–3734.

6. Zenke F, Gerstner W, Ganguli S: **The temporal paradox of Hebbian learning and homeostatic plasticity**. *Current Opinion in Neurobiology* 2017, **43**:166–176.

7. Tien N-W, Kerschensteiner D: **Homeostatic plasticity in neural development**. *Neural Development* 2018, **13**.

8. Debanne D, Inglebert Y, Russier M: **Plasticity of intrinsic neuronal excitability**. *Current Opinion in Neurobiology* 2019, **54**:73–82.

9. Swensen AM, Bean BP: **Robustness of Burst Firing in Dissociated Purkinje Neurons with Acute or Long-Term Reductions in Sodium Conductance**. *Journal of Neuroscience* 2005, **25**:3509–3520.

10. He LS, Rue MCP, Morozova EO, Powell DJ, James EJ, Kar M, Marder E: **Rapid adaptation to elevated extracellular potassium in the pyloric circuit of the crab, Cancer borealis**. *Journal of Neurophysiology* 2020, **123**:2075–2089.

11. Lane BJ, Samarth P, Ransdell JL, Nair SS, Schulz DJ: **Synergistic plasticity of intrinsic conductance and electrical coupling restores synchrony in an intact motor network**. *eLife* 2016, **5**:e16879.

12. Attwell D, Laughlin SB: **An Energy Budget for Signaling in the Grey Matter of the Brain**. *Journal of Cerebral Blood Flow & Metabolism* 2001, **21**:1133–1145.

13. Sengupta B, Stemmler M, Laughlin SB, Niven JE: **Action Potential Energy Efficiency Varies Among Neuron Types in Vertebrates and Invertebrates**. *PLoS Comput Biol* 2010, **6**:e1000840.

14. Stemmler M, Koch C: **How voltage-dependent conductances can adapt to maximize the information encoded by neuronal firing rate**. *Nat Neurosci* 1999, **2**:521–527.

15. Vergara RC, Jaramillo-Riveri S, Luarte A, Moënne-Loccoz C, Fuentes R, Couve A, Maldonado PE: **The Energy Homeostasis Principle: Neuronal Energy Regulation Drives Local Network Dynamics Generating Behavior**. *Front Comput Neurosci* 2019, **13**:49.

16. Remme MWH, Rinzel J, Schreiber S: **Function and energy consumption constrain neuronal biophysics in a canonical computation: Coincidence detection**. *PLOS Computational Biology* 2018, **14**:e1006612.

17. Adams S, Zubov T, Bueschke N, Santin JM: **Neuromodulation or energy failure? Metabolic limitations silence network output in the hypoxic amphibian brainstem**. *American Journal of Physiology-Regulatory, Integrative and Comparative Physiology* 2021, **320**:R105–R116.

18. Sætra MJ, Einevoll GT, Halnes G: **An electrodiffusive, ion conserving Pinsky-Rinzel model with homeostatic mechanisms**. *PLoS Comput Biol* 2020, **16**:e1007661.

19. O'Leary T, Williams AH, Franci A, Marder E: **Cell Types, Network Homeostasis, and Pathological Compensation from a Biologically Plausible Ion Channel Expression Model**. *Neuron* 2014, **82**:809–821.

20. Joseph A, Turrigiano GG: **All for One But Not One for All: Excitatory Synaptic Scaling and Intrinsic Excitability Are Coregulated by CaMKIV, Whereas Inhibitory Synaptic Scaling Is Under Independent Control**. *The Journal of Neuroscience* 2017, **37**:6778–6785.

21. Yang J, Shakil H, Prescott SA: **Simultaneously regulating many properties requires that neurons adjust diverse ion channels**. *bioRxiv* 2020, doi:10.1101/2020.12.04.410787.

22. Tyssowski KM, Letai KC, Rendall SD, Tan C, Nizhnik A, Kaeser PS, Gray JM: **Firing Rate Homeostasis Can Occur in the Absence of Neuronal Activity-Regulated Transcription**. *J Neurosci* 2019, **39**:9885–9899.

23. Jamann N, Dannehl D, Lehmann N, Wagener R, Thielemann C, Schultz C, Staiger J, Kole MHP, Engelhardt M: **Sensory input drives rapid homeostatic scaling of the axon initial segment in mouse barrel cortex**. *Nat Commun* 2021, **12**:23.

24. Hesse J, Schreiber S: **Externalization of neuronal somata as an evolutionary strategy for energy economization**. *Current Biology* 2015, **25**:R324–R325.

25. Samengo I, Mato G, Elijah DH, Schreiber S, Montemurro MA: **Linking dynamical and functional properties of intrinsically bursting neurons**. *Journal of Computational Neuroscience* 2013, **35**:213–230.

26. Blankenburg S, Wu W, Lindner B, Schreiber S: **Information filtering in resonant neurons**. *Journal of Computational Neuroscience* 2015, **39**:349–370.

27. Ratté S, Hong S, De Schutter E, Prescott SA: **Impact of neuronal properties on network coding: roles of spike initiation dynamics and robust synchrony transfer**. *Neuron* 2013, **78**:758–772.

28. Hesse J: **Implications of neuronal excitability and morphology for spike-based information transmission**. 2017, doi:10.18452/18583.

29. Gjorgjieva J, Mease RA, Moody WJ, Fairhall AL: **Intrinsic Neuronal Properties Switch the Mode of Information Transmission in Networks**. *PLoS Computational Biology* 2014, **10**:e1003962.

30. Smeal RM, Ermentrout GB, White JA: **Phase-response curves and synchronized neural networks**. *Phil Trans R Soc B* 2010, **365**:2407–2422.

31. Schleimer J-H, Stemmler M: **Coding of Information in Limit Cycle Oscillators**. *Phys Rev Lett* 2009, **103**:248105.

32. Hodgkin AL: **The local electric changes associated with repetitive action in a non-medullated axon**. *The Journal of Physiology* 1948, **107**:165–181.

33. Prescott SA, De Koninck Y, Sejnowski TJ: **Biophysical Basis for Three Distinct Dynamical Mechanisms of Action Potential Initiation**. *PLoS Comput Biol* 2008, **4**:e1000198.

34. Kirst C: **Synchronization, Neuronal Excitability, and Information Flow in Networks of Neuronal Oscillators**. 2012,

35. Izhikevich EM: *Dynamical systems in neuroscience: the geometry of excitability and bursting*. MIT Press; 2007.

36. Guckenheimer J, Labouriau JS: **Bifurcation of the Hodgkin and Huxley equations: A new twist**. *Bulletin of Mathematical Biology* 1993, **55**:937–952.

37. Stiefel KM, Gutkin BS, Sejnowski TJ: **The effects of cholinergic neuromodulation on neuronal phase-response curves of modeled cortical neurons**. *J Comput Neurosci* 2009, **26**:289–301.

38. Stiefel KM, Gutkin BS, Sejnowski TJ: **Cholinergic Neuromodulation Changes Phase Response Curve Shape and Type in Cortical Pyramidal Neurons**. *PLoS ONE* 2008, **3**:e3947.

39. Zhao Z, Gu H: **Transitions between classes of neuronal excitability and bifurcations induced by autapse**. *Sci Rep* 2017, **7**:6760.

40. Morozova EO, Zakharov D, Gutkin BS, Lapish CC, Kuznetsov A: **Dopamine Neurons Change the Type of Excitability in Response to Stimuli**. *PLOS Computational Biology* 2016, **12**:e1005233.

41. Franci A, Drion G, Seutin V, Sepulchre R: **A Balance Equation Determines a Switch in Neuronal Excitability**. *PLoS Comput Biol* 2013, **9**:e1003040.

42. Hesse J, Schleimer J-H, Schreiber S: **Qualitative changes in phase-response curve and synchronization at the saddle-node-loop bifurcation**. *Phys Rev E* 2017, **95**:052203.

43. Contreras SA, Schleimer J-H, Gulledge AT, Schreiber S: **Activity-mediated accumulation of potassium induces a switch in firing pattern and neuronal excitability type**. *PLOS Computational Biology* 2021, **17**:e1008510.

44. Gutkin BS, Ermentrout GB: **Dynamics of Membrane Excitability Determine Interspike Interval Variability: A Link Between Spike Generation Mechanisms and Cortical Spike Train Statistics**. *Neural Computation* 1998, **10**:1047–1065.

45. Zhuchkova E, Remme MWH, Schreiber S: **Somatic versus Dendritic Resonance: Differential Filtering of Inputs through Non-Uniform Distributions of Active Conductances**. *PLoS ONE* 2013, **8**:e78908.

46. Rau F, Clemens J, Naumov V, Hennig RM, Schreiber S: **Firing-rate resonances in the peripheral auditory system of the cricket, Gryllus bimaculatus**. *Journal of Comparative Physiology A* 2015, **201**:1075–1090.

47. Ermentrout GB, Galán RF, Urban NN: **Relating Neural Dynamics to Neural Coding**. *Physical Review Letters* 2007, **99**.

48. Galán RF, Bard Ermentrout G, Urban NN: **Reliability and stochastic synchronization in type I vs. type II neural oscillators**. *Neurocomputing* 2007, **70**:2102–2106.

49. Kirst C, Ammer J, Felmy F, Herz A, Stemmler M: **Fundamental Structure and Modulation of Neuronal Excitability: Synaptic Control of Coding, Resonance, and Network Synchronization**. 2015, doi:10.1101/022475.

50. Dumortier F, Roussarie R, Sotomayor J, Zoladek H: *Bifurcations of Planar Vector Fields: Nilpotent Singularities and Abelian Integrals*. Springer-Verlag; 1991.

51. Franci A, O'Leary T, Golowasch J: **Positive Dynamical Networks in Neuronal Regulation: How Tunable Variability Coexists With Robustness**. *IEEE Control Syst Lett* 2020, **4**:946–951.

52. Roemschied FA, Eberhard MJ, Schleimer J-H, Ronacher B, Schreiber S: **Cell-intrinsic mechanisms of temperature compensation in a grasshopper sensory receptor neuron**. *eLife* 2014, **3**:e02078.

53. O'Leary T: **Homeostasis, failure of homeostasis and degenerate ion channel regulation**. *Current Opinion in Physiology* 2018, **2**:129–138.

54. Ding L, Hou C: **Stabilizing control of Hopf bifurcation in the Hodgkin–Huxley model via washout filter with linear control term**. *Nonlinear Dyn* 2010, **60**:131–139.

55. Lu J, Engl HW, Schuster P: **Inverse bifurcation analysis: application to simple gene systems**. *Algorithms Mol Biol* 2006, **1**:11.

56. Dobson I: **Distance to Bifurcation in Multidimensional Parameter Space: Margin Sensitivity and Closest Bifurcations**. In *Bifurcation Control*. Edited by Chen G, Hill DJ, Yu X. Springer Berlin Heidelberg; 2004:49–66.

57. Carrillo FA, Verduzco F: **Control of the Planar Takens–Bogdanov Bifurcation with Applications**. *Acta Appl Math* 2009, **105**:199–225.

58. de Carvalho Braga D, Fernando Mello L, Rocşoreanu C, Sterpu M: **Control of planar Bautin bifurcation**. *Nonlinear Dyn* 2010, **62**:989–1000.

59. Wei Kang, Ke Liang: **The stability and invariants of control systems with pitchfork or cusp bifurcations**. In *Proceedings of the 36th IEEE Conference on Decision and Control.* . 1997:378–383 vol.1.

60. Huang C, Sun W, Zheng Z, Lu J, Chen S: **Hopf bifurcation control of the M–L neuron model with type I**. *Nonlinear Dyn* 2017, **87**:755–766.

61. Crambert G, Hasler U, Beggah AT, Yu C, Modyanov NN, Horisberger J-D, Lelièvre L, Geering K: **Transport and Pharmacological Properties of Nine Different Human Na,K-ATPase Isozymes \***. *Journal of Biological Chemistry* 2000, **275**:1976–1986.

62. Gorur-Shandilya S, Marder E, O'Leary T: **Activity-dependent compensation of cell size is vulnerable to targeted deletion of ion channels**. *Sci Rep* 2020, **10**:15989.

63. Kulik Y, Jones R, Moughamian AJ, Whippen J, Davis GW: **Dual separable feedback systems govern firing rate homeostasis**. *eLife* 2019, **8**:e45717.

64. O'Leary T, Marder E: **Temperature-Robust Neural Function from Activity-Dependent Ion Channel Regulation**. *Current Biology* 2016, **26**:2935–2941.

65. Schleimer J-H, Schreiber S: **Homeostasis: How Neurons Achieve Temperature Invariance**. *Current Biology* 2016, **26**:R1141–R1143.

66. Schleimer J-H, Hesse J, Contreras SA, Schreiber S: **Firing statistics in the bistable regime of neurons with homoclinic spike generation**. *Phys Rev E* 2021,

67. Boulware MJ, Marchant JS: **Timing in Cellular Ca2+ Signaling**. *Current Biology* 2008, **18**:R769–R776.

68. Triesch J: **Synergies Between Intrinsic and Synaptic Plasticity Mechanisms**. *Neural Computation* 2007, **19**:885–909.

69. Cannon J, Miller P: **Synaptic and intrinsic homeostasis cooperate to optimize single neuron response properties and tune integrator circuits**. *Journal of Neurophysiology* 2016, **116**:2004–2022.

70. Cannon J, Miller P: **Stable Control of Firing Rate Mean and Variance by Dual Homeostatic Mechanisms**. *J Math Neurosc* 2017, **7**:1.

71. Miller P, Cannon J: **Combined mechanisms of neural firing rate homeostasis**. *Biol Cybern* 2019, **113**:47–59.

72. Drion G, O'Leary T, Marder E: **Ion channel degeneracy enables robust and tunable neuronal firing rates**. *Proc Natl Acad Sci USA* 2015, **112**:E5361–E5370.

73. Transtrum MK, Machta BB, Brown KS, Daniels BC, Myers CR, Sethna JP: **Perspective: Sloppiness and emergent theories in physics, biology, and beyond**. *The Journal of Chemical Physics* 2015, **143**:010901.

74. Gutenkunst RN, Waterfall JJ, Casey FP, Brown KS, Myers CR, Sethna JP: **Universally Sloppy Parameter Sensitivities in Systems Biology Models**. *PLoS Computational Biology* 2007, **3**:e189.

75. OLSHAUSEN B, FIELD D: **Sparse coding of sensory inputs**. *Current Opinion in Neurobiology* 2004, **14**:481–487.

76. Chalk M, Marre O, Tkačik G: **Toward a unified theory of efficient, predictive, and sparse coding**. *Proc Natl Acad Sci USA* 2018, **115**:186–191.

77. Gulledge AT, Dasari S, Onoue K, Stephens EK, Hasse JM, Avesar D: **A Sodium-Pump-Mediated Afterhyperpolarization in Pyramidal Neurons**. *Journal of Neuroscience* 2013, **33**:13025–13041.

78. Arganda S, Guantes R, de Polavieja GG: **Sodium pumps adapt spike bursting to stimulus statistics**. *Nat Neurosci* 2007, **10**:1467–1473.

79. Allen NJ, Lyons DA: **Glia as architects of central nervous system formation and function**. *Science* 2018, **362**:181–185.

80. Bianchi D, Marasco A, Limongiello A, Marchetti C, Marie H, Tirozzi B, Migliore M: **On the mechanisms underlying the depolarization block in the spiking dynamics of CA1 pyramidal neurons**. *J Comput Neurosci* 2012, **33**:207–225.

81. Newby AC: **Adenosine and the concept of 'retaliatory metabolites.'** *Trends in Biochemical Sciences* 1984, **9**:42–44.

82. Giménez-Cassina A, Martínez-François JR, Fisher JK, Szlyk B, Polak K, Wiwczar J, Tanner GR, Lutas A, Yellen G, Danial NN: **BAD-Dependent Regulation of Fuel Metabolism and KATP Channel Activity Confers Resistance to Epileptic Seizures**. *Neuron* 2012, **74**:719–730.

83. Boison D, Steinhäuser C: **Epilepsy and astrocyte energy metabolism**. *Glia* 2018, **66**:1235–1243.

84. Venkatesh KV, Bhartiya S, Ruhela A: **Multiple feedback loops are key to a robust dynamic performance of tryptophan regulation in *Escherichia coli***. *FEBS Letters* 2004, **563**:234–240.